\documentclass{elsart1p}
\usepackage{graphicx}
\usepackage{amssymb}

\def\lsim{\mathrel{\rlap{\lower3pt\hbox{\hskip1pt$\sim$}}
     \raise1pt\hbox{$<$}}} 
\def\gsim{\mathrel{\rlap{\lower3pt\hbox{\hskip1pt$\sim$}}
     \raise1pt\hbox{$>$}}}

\def\be{\begin{eqnarray}}\def\ba{\begin{eqnarray}}
\def\ee{\end{eqnarray}}\def\ea{\end{eqnarray}}
\def\ben{\begin{enumerate}}\def\bitem{\begin{itemize}}
\def\een{\end{enumerate}}\def\eitem{\end{itemize}}

\begin{document}
\begin{frontmatter}
%
%
%
\title{Weak Interaction in Holographic QCD}
%
%
\author[dg]{Doron Gazit}
%
\author[huy]{Ho-Ung Yee}
\address[dg]{Institute for Nuclear Theory, University of Washington, Box 351550, 98195, Seattle, WA, USA.}
\address[huy]{Abdus Salam International Centre for Theoretical Physics, Strada Costiera 11, 34014 Trieste, Italy.}
\begin{abstract}
We present a recent proposal of a simple prescription for including low-energy
weak-interactions into the framework of holographic QCD, based on
the standard AdS/CFT dictionary of double-trace deformations. This opens a new perspective on
phenomenological applications of holographic QCD: calculating weak observables in the strongly coupled regime of QCD. The idea is
general enough to be implemented in any holographic model. Its efficiency and usefulness are demonstrated by performing few exemplar calculations of weak reactions involving mesons and baryons, within the Sakai-Sugimoto and hard/soft wall holographic models.
\end{abstract}
\begin{keyword}
\PACS
\end{keyword}
\end{frontmatter}
Large N factorization of gauge invariant master fields, as well as renormalization group flow, in the large N limit of QCD, suggest the existence of a classical theory in five dimensions dual to large N QCD, named holographic QCD, with the fifth dimension roughly corresponding to the energy scale. In view of the known AdS/CFT correspondence \cite{1}, we expect that the would-be 5-dimensional holographic dual
theory becomes highly non-local in UV regime where
the corresponding large N QCD is asymptotically free, and a local theory in IR region as large N QCD
becomes strongly coupled.

The existing holographic QCD models, like a top-down Sakai-Sugimoto
model \cite{2} or a bottom-up Hard/Soft Wall model
\cite{3}, effectively model the strongly coupled regime using a local theory.
They capture important aspects
of low energy QCD such as chiral symmetry and confinement, and
explain related experimental observables up to 20\%, which one might
expect from large N approximation. The previous calculations in this
framework concern mostly the pure QCD sector and its electromagnetic
couplings. As there are many important processes which
involve both weak-interaction and strongly-coupled low energy QCD,
it is pertinent to include effective weak-interactions in the
framework of holographic QCD, which will allow us a new tool for
estimating various hadronic weak processes.

The weak interactions are mediated by exchanges of heavy $W^{\pm}$
or $Z^0$ bosons. The tree level weak interactions, in the low energy
limit, in which $q \ll M_{W^{\pm},Z^0}$, become
effective Fermi point-like vertices. We present here a recent proposal for a simple
prescription for introducing these effective Fermi vertices of
weak-interaction in holographic QCD \cite{4}. In view of the original QCD,
this corresponds to perturbing the Lagrangian by the effective
four-fermi operators
\begin{eqnarray} \label{weakvertex}
{\Delta {\cal L}_{weak}} &=&{{4 G_F \over
\sqrt{2}}\left(J_{W^+} J_{W^-}
+\cos^2\theta_W\left(J_{Z^0}\right)^2\right)=} \\ \nonumber &=& {4 G_F \over
\sqrt{2}}\left((J^1_L)^2+(J^2_L)^2+ \left(J^3_L-\sin^2\theta_W
J_Q\right)^2\right)
\end{eqnarray}
where
$J_L^a=\sum_f\bar\psi^f_L{\sigma^a\over 2}\psi^f_L$ and $J_Q$ are
$SU(2)_L$ and electromagnetic currents respectively, and $\theta_W$
is the weak angle. The currents in the above
may include leptonic part as well, treated as an
external background.

In holographic QCD as a gauge/gravity correspondence, perturbing the QCD Lagrangian would correspond to
deforming boundary conditions of 5D fields near the UV boundary in a suitable way. The weak perturbation in
Eq.~(\ref{weakvertex}), is quadratic in the currents, which are single-trace operators. According to
Ref.~\cite{5}, each current in this deformation corresponds to a 5D gauge
 field $A_\mu(x,r)$, where $r$ is the
5'th coordinate, that behaves near the UV boundary $r\to\infty$ as $ A_\mu(x,r)\sim c_1 r^{-\Delta_-}+c_2
r^{-\Delta_+}$, where a non-normalizable coefficient
$c_1$ is equal to the functional derivative of the weak interaction of Eq.~(\ref{weakvertex})
with
respect to a current, and $c_2$ corresponds to the expectation value of
the current, a
normalizable mode.

This is the essence of our proposal, as it defines the boundary conditions of
the 5D field. The equation of motion itself is derived
from the holographic QCD model. These two ingredients are sufficient
to obtain the behavior of the field in the entire 5D space, including at the IR,
where its value corresponds to the low-energy limit of QCD, hence to the physical
value of the current corresponding to the field, according to the holographic model.

Although the idea is general enough to be applicable to any model of
holographic QCD (see Ref.\cite{6} for a review), we use two examples, the Sakai-Sugimoto model  and the Hard/Soft Wall model, to illustrate our proposal. In order to show
its usefulness, we present the main achievements of few exemplar physical calculations.

{\it Charged pion decay $\pi^+ \to \mu^+ \nu_\mu$} -- In order to incorporate pions, we use the usual unitary gauge for the 5th coordinate gauge field. For example, in the Sagai-Sugimoto model we take $A_Z=\frac{2}{\pi f_\pi} \frac{1}{1+Z^2}\pi(x)$. Together with our prescription, that connects the gauge field at the UV with the external muon current, this leads to an interaction term in the Lagrangian, which one quickly identifies as the result achieved by current algebra, ${\cal{L}}=_{\pi\bar l \nu}=-2 G_F f_\pi (\partial_\mu \pi^-)(\bar l_L \gamma^\mu \nu_L)$.

{\it Neutron $\beta$-decay} -- In the use of our prescription,
this case is identical to the pion decay, as it includes a coupling of
the neutron to an external current. However,
one needs a specific approach for the nucleon itself.
This can be done by using an effective instanton model for the nucleon,
calibrated to give the nucleon mass, which interacts with the gauge field through covariant derivatives and magnetic moments \cite{7}. In the case of the Sagai-Sugimoto model, this constraint is sufficient to remove all the freedom in the calculation, and leads to an amazing prediction for the axial constant $g_A  \approx 1.30$, compared with the experimental axial constant extracted from neutron-decay $g_A=1.2695(29)$. This reflects an under-prediction of about 4\% in the half-life of the neutron. In the hard wall model, an additional parameter is needed when describing the nucleon, due to the coupling to the bi-fundamental field, by which the chiral symmetry is modeled. Thus, in this case we can only relate between this parameter $D$ and the axial constant $g_A=(0.33+1.02 D)$ (D is measured in GeV$^{-1}$).

{\it Parity Non-conserving (PNC) pion-nucleon interaction} -- Our final example has been
difficult to obtain with other conventional tools. It is also a
first non-trivial example which does not involve external leptons,
so that the full aspect of our proposal should be used for its
calculation.
We are interested in the parity-violating couplings of mesons,
especially the pions, to the nucleons, induced by weak-interactions.
For an illustrative purpose, we focus here only on the charged
pion-nucleon coupling from a $W^\pm$-exchange. It is the pion in our case which carries an induced external source, which then couples to the nucleon in the same way the electron and anti-neutrino couple to the nucleon in the neutron decay.
For the $q^2\to 0$ limit we can write the PNC interaction ${\cal L}^{weak}_{ N-\pi}=-2 G_F f_\pi (\bar
p\gamma^\mu n) \left(\partial_\mu\pi^+\right)$. This term is identical to
the contribution of pion exchange to the PNC coupling, as achieved
by current algebra and chiral perturbation theory \cite{8}.
Although the above pion-nucleon coupling in $q^2\to 0$ limit is
largely dictated by chiral algebra alone, its non-trivial
$q^2$-dependence when we consider finite $q^2$ is beyond the ability
of the chiral algebra. Our framework can give predictions for that.
Our method can also provide predictions for the coupling of other
excited mesons such as $\rho$, $\omega$, {\it etc}, whose details
will be reported soon elsewhere \cite{9}.

To summarize, we have outlined a proposal for including effective weak
interactions in the framework of holographic QCD, and presented its application to three low energy processes, in two holographic models, the Sakai-Sugimoto and Hard/Soft Wall
models. Though these examples are only for small, practically zero, momentum transfer,
the prescription we propose should be valid to
moderately higher energies before QCD asymptotic freedom sets in, enabling calculations in energies which are not assessable in other methods.
Contrary to other theoretical tools to calculate such reactions,
e.g. chiral perturbation theory, the current approach not only
recovers the operator structure of previous methods, but also gives
a quantitative estimate, up to about 10-20\%, of the coupling
constants.

This work was supported, in part, by DOE grant number DE-FG0200ER41132.

\end{document}